\documentclass[11pt]{article}
\pdfoutput=1

\usepackage{fullpage}
\usepackage{algorithm}
\usepackage{algorithmic}

\usepackage[numbers]{natbib}
\usepackage{hyperref}
\usepackage{amsthm}
\usepackage{amsmath}
\usepackage{amsfonts}
\usepackage{breqn}
\usepackage{graphicx}
\usepackage{subfig}
\usepackage[table]{xcolor}
\usepackage{multirow}

\theoremstyle{definition}

\theoremstyle{remark}

\begin{document}

\title{PeRView: A Framework for Personalized Review Selection Using Micro-Reviews}
\author{Muhmmad Al-Khiza'ay 1\thanks{Author 1's School of Information Technology, Deakin University, 221 Burwood Highway, Vic 3125, Australia. \texttt{malkhiza@@deakin.edu.au}.}
\and Noora Alallaq 2\thanks{Author 2's School of Information Technology, Deakin University, 221 Burwood Highway, Vic 3125, Australia. \texttt{nalallaq@deakin.edu.au}.}
\and Qusay Alanoz 3\thanks{Author 3's Faculty of Mathematics and Computer Science, University of Lodz, Lodz, Poland. \texttt{qusay.arif2015@gmail.com}.}
\and Adil Al-Azzawi 4\thanks{Author 4's Electrical Engineering and Computer Science Dept. University of Missouri-Columbia Missouri, Columbia, USA. \texttt{aaadn5@mail.missouri.edu}.}
\and N.Maheswari 5\thanks{Author 5's School of Computing Science and Engineering, Vellore institute of technology, Chennai, India. \texttt{maheswari.n@vit.ac.in}.}
}

\maketitle
\thispagestyle{empty}

\begin{abstract}
In the contemporary era,
social media has its influence on people in making decisions.
The proliferation of online reviews with diversified
and verbose content often causes problems inaccurate decision making.
Since online reviews have an impact on people
of all walks of life while taking decisions,
choosing appropriate reviews based on
the podsolization consisting
is very important since it relies
on using such micro-reviews consistency
to evaluate the review set section.
Micro-reviews are very concise and directly
talk about product or service instead of
having unnecessary verbose content.
Thus, micro-reviews can help in
choosing reviews based on their
personalized consistency that is
related to directly or indirectly to
the main profile of the reviews.
Personalized reviews selection that
is highly relevant with high personalized
coverage in terms of matching with micro-reviews
is the main problem that is considered in this paper.
Furthermore, personalization with user preferences
while making review selection is also considered
based on the personalized users' profile.
Towards this end, we proposed a framework known
as PeRView for personalized review selection using
micro-reviews based on the proposed evaluation metric
approach which considering two main factors
(personalized matching score and subset size).
Personalized Review Selection Algorithm (PRSA)
is proposed which makes use of multiple
similarity measures merged to have highly
efficient personalized reviews matching
function for selection.
The experimental results based on
using reviews dataset which is
collected from (\url{www.YELP.COM})
while micro-reviews dataset is obtained from
(\url{www.Foursquare.com}).
show that the personalized reviews selection
is a very empirical case of study.
\end{abstract}

\newpage
\thispagestyle{empty}

\tableofcontents

\newpage
\pagenumbering{arabic}




\section{Introduction} \label{sec-intro}

Online reviews for almost any product or service
have been influencing the people's decision making process.
A reviewer basically produces an assessment about a company, service,
or any other task~Vasconcelos~{et~al.}~\cite{vasconcelos2014makes}.
In these days,
we can find a variety
of ample reviewers in different web sources.
For instance,
in online shopping,
web services such as (\url{http://Amazon.com})
provide hosts reviews as a part of the online shopping experience.
This part provides a clearly assessment
and product ranking that helps and
assist their and other customers in determining
which product is the most suitable for their need~Agarwal~{et~al.}~\cite{agarwal2011towards}.
Moreover,
this is the reason that many businesses are focusing on
getting better reviews to promote their businesses
based on the best reviews that are personally considering
such similar consistency~\cite{Nguyen2015}~\cite{NguyenLT13}.

Intuitively,
customers are more inundated by
the variety of comments that made during numerous reviews.
In this case,
it is not clearly enough
which review comments are worth or not for the reader's attention,
and this is worsened by two factors,
the length and the verbosity of many reviews.
It is more increasingly that is more difficult
to discover and determine the personalized review
that has a higher relevant experience~Bodke~{et~al.}~\cite{bodke2015survey}.
Furthermore,
high personalized quality reviews identification
and selection is a hard task for the customers
which is potential topic that has been
more focusing of substantial research ~Chen~{et~al.}~\cite{chen2011quality}.

Recently,
social networking has been growing intensively
and has been became more established from year-to-year.
In this case,
the emergence of new type of online reviews
has been discovered which called micro-reviews service~Nguyen~{et~al.}~\cite{NguyenLT15}.
Micro-review is prevailing since the emerging trends
in social media and micro-blogging which
is alternative source of information for the reviewers to
find more interested information to read.
Micro-reviews are consisting, short, and focused,
as well as they are nicely complementing,
elaborate, and verbose nature of full-text reviews focusing
on a specific of an item~Chong~{et~al.}~\cite{chong2015did}.
Moreover,
the micro-review cannot be properly expanded
more than 140 characters which means that
the reviewers should be focusing on aspect of the venue
that are more important to the investigators (users)~\cite{kudyba2014big}.

(\url{www.YELP.COM}) and
(\url{www.Foursquare.com}) are popular social web
sites that is in the meanwhile most customers used,
provide online reviews section for public.
For instance,
(\url{www.YELP.COM}) provides reviews on restaurants
while (\url{www.Foursquare.com}) provides micro-reviews
that are more concise and focused on the feedback.
People who wanted to visit a restaurant and before making a decision,
generally find reviews in these social web sites
where these reviews are related to restaurants across the globe.
Many researchers contributed
towards analysis of online reviews~\cite{GhoseI07,KimPCP06,lappas2010efficient,Nguyen2015}.

In this paper,
we present an approach to selecting the personalized reviews
that are highly relevant with high personalized coverage
in terms of matching with a selecting sub-set of micro reviews.
The selected sub-set contains more focused and concise
by reflecting the true opinion of individuals.
Moreover,
selected sub-set of micro-reviews are not verbose and do not contain unnecessary information.
The idea is to select a personalized subset of reviews,
pertaining to any restaurant or service,
that exhibit high personalized coverage in terms of matching with micro-reviews sub-set.
For the purpose of matching it is essential
to have text mining approaches and lexical dictionary
such as \textit{WordNet} to have comprehensive matching strategy.
Our matching strategy includes different personalized matching criteria
such as syntactical, semantic and sentiment matching.
These diversified personalized matching approaches are merged
to have a significant personalized matching function
based on two factors (personalized matching score and the reviews selected size)
that can be used to have high satisfactory reviews.
Towards this end, we proposed a framework named \textit{PeRView}
with an underling algorithm named Personalized Review Selection Algorithm (\textit{PRSA})
which takes micro-reviews and reviews as input and produce selected reviews.
\textit{PRSA} employs the matching function aforementioned and personalized preferences
as well for effective review selection process.

In this paper, the real time online reviews of restaurants
and corresponding micro-reviews are collected from
two different datasets~(\url{www.YELP.COM}) and~(\url{www.Foursquare.com}) respectively.
These reviews and micro-reviews for more than 50 restaurants
formed the dataset for experiments. In this case,
the high quality personalized preview selection is proposed
to select a subset of the personalized reviews
that are highly correlated and related to the selected micro-reviews sub-set.
The first contribution in this paper is that proposing
a micro-review sub-set selection instead of relying on the whole set of the micro-review.
The selected sub-set of micro-review's algorithm based
on select the set of micro-reviews that covers the whole domain.
Then, the second contribution in this research is
on the a framework named \textit{PeRView}
which facilitates online personalized review selection.
The framework also supports personalization
by considering preferences of end users
by building a personal profile for each user (reviewers).
Then, based on the third contribution of this paper
which design a personalized evaluation matric of review's selection,
the highly personalized reviews set that coverage significant consistence
of the selected micro-reviews are selected
based on considering two main factors
(average personalized matching score and sub-set size).

The rest of the paper is structured as follows.
Section~\ref{sec-realted-works} related works and preliminary.
Section~\ref{sec-proposed-methodology} presents the proposed methodology.
Section~\ref{sec-exp-results} presents experimental results.
Section~\ref{sec-xyz}
throws light into conclusions drawn
and recommendations for future research in the area of review selection.

\section{Related Works and Preliminary} \label{sec-realted-works}
\subsection{Related works}

This section reviews important literature related
to online review selection.
~\cite{ChienC08}employed selection concept with data mining
concepts but it was meant for personnel selection.
They built a framework for selection of personnel.
Meng~{et~al.}~\cite{MengDZC14},
proposed a recommendation method known
as Keyword-Aware Service Recommendation (KASR).
They used personalized recommendations where
personalized concept has similarities of the work of this paper.
User preferences are used for personalized ratings.
Bhatia~{et~al.}~\cite{bhatia2016novel},
proposed a method for review selection process using micro-reviews.
Bhatia et al believed that micro reviews
are concise and they are best used for
validating reviews and choose good reviews
that match with micro-reviews as much as possible.
Our work is close to this work but we improved
it further and also personalized.
Nie~{et~al.}~\cite{nie20173d} employed LDA for music,
images and text. As far as LDA is concerned,
our approach in this paper is reusing
LDA to get an improved form
for review selection process.
~\cite{Hong010}
focused on making topic models in Twitter.
They employed LDA for empirical study
on topic modelling in Twitter.
~\cite{risch2016detecting} on the other hand
studied and used LDA for detecting Twitter topics.

Blei~{et~al.}~\cite{BleiNJ03} investigated
the utility of LDA for topic modelling.
As topic modelling can reveal explicit details of a document,
this method was used for text modelling
and document classification. Mason et al.
~\cite{MasonGDCHDTM16} employed LDA for
micro-summarization of online reviews to
render useful information to end users.
They used unsupervised and supervised methods
for empirical study.
They employed multiple aspects in the
research such as entity recognition,
summarization and sentiment analysis.
lappas~{et~al.}~\cite{LappasCT12} focused on review selection.
Their work was meant for filling the gap
between review summarization and review selection processes.
They considered it as combinatorial optimization problem.
Tsaparas~{et~al.}~\cite{tsaparas2011selecting}
studied user reviews made online and explored
a method for selecting a comprehensive collection
of reviews that make sense.
Various complexity coverage problems are addressed analyzing them.

Ganesan~{et~al.}~\cite{GanesanZV12} made it explored
an unsupervised approach for generating
summary of opinions. They proposed
a methodology for generating ultra-concise
summaries of sentiments.
They achieved it by using some heuristic algorithms.
Lu~{et~al.}~\cite{LuMPT16} opined that micro-reviews
are concise and more meaningful than reviews.
They studied probabilistic models based on
LDA for topic suggestion for micro-reviews.
This approach is based on sentiments.
Nguyen~{et~al.}~\cite{NguyenLT13} also used micro-reviews
in order to have efficient selection of reviews.
They used micro-reviews to obtain salient features
of reviews and finally select best reviews.
This work is also similar to our work in this paper.
However it follows crowdsourced approach.
Similar work is found in Nguyen~{et~al.}~\cite{NguyenLT13}
where micro-reviews are used to test coverage and choose reviews.
Selvam~\cite{selvam2016unified}
employed Integer Linear Programming (ILP)
and unified classification for review selection with the help of micro-reviews.

Nguyen~{et~al.}~\cite{NguyenLT17} continued
their research on micro-reviews.
They explored review synthesis for summarization
of micro-reviews for making them more compact and readable.
Nguyen~{et~al.}~\cite{Nguyen2015} studied
mining of massive textual data in order to obtain
heuristics for achieving selection problem.
They proposed a subset selection procedure by
extending Neyman-Pearson feature selection approach.
More on feature selection algorithms is
found in ~\cite{mukhopadhyay2014survey},
~\cite{XueZBY16} and~\cite{ChandrashekarS14}.
Liu~{et~al.}~\cite{LiuMSZ10a}[27] on the other hand
used feature selection approach used in
data mining and its utility in reducing
complexity in data mining procedures.
Lu~{et~al.}~\cite{LuZS09} used
short comments and investigated the problem
of rated aspect summarization of target entity
to have knowledge on such data.
lappas~{et~al.}~\cite{KangLTLCX12} proposed
a method for finding top-k web service recommendations
where selection of recommendations is based on
a hybrid similarity metric.
Adomavicius~{et~al.}~\cite{adomavicius2015multi} explored user
modelling with multi-criteria for generating recommendations.
Chorley~{et~al.}~\cite{chorley2013visiting} studied
the users of online review web site Foursquare.com.
They focused on finding personality of users based on
their visiting patterns and other content based analysis.
Based on the reviews made by users,
they tried to estimate different aspects of users' personality.

Chen~{et~al.}~\cite{ChenYHZ16} studied
user generated content (UGC) available in Foursquare.com
for investigating diversity of tips, venue categories,
and measuring of tips for sentiment analysis.
Tao~{et~al.}~\cite{TaoLZ11} focused on web
information gathering by using personalized ontology.
Personalized data of end users is used for information gathering.
In the literature many approaches are found for review selection.
However, they used different approaches for review selection.
Some of them used micro-reviews for finding coverage
of reviews and select them.
In this paper we improve review selection
based on micro-review approaches based
on matching score ranking using Mutual Information (MI)
feature ranking approach which helps
the proposed system to avoid the similar matching score that intuitively accrue in some cases.\\

\subsection{Preliminary}
Review is an evaluation of a publication or product or service.
Reviews have influence on quick
understanding and sometimes
even making decisions as well.
It may reflect relative merit
of publication or service.
Consumers of a product or
service may write a review on it.
Review which is denoted as \textit{R}
generally contains more
information while micro-review
has limited and concise information.
In this case, we have to
define some relational terms
in term to fully understand
the main problem. Literary,
reviews and micro-reviews
comments are a set of words
where the single word mathematically
denoted as \textit{W}. Although,
the review is a group of
sentences that are part of a
review denoted as the following equation~\ref{equ-review}:

\begin{equation} \label{equ-review}
R_{R\in\mathbb{R}}=\{s_1, s_2, ... ,s_n\}=\sum_{i=1}^{R}S_i
\end{equation}
Where s is denoted as
a sentence and it is
define as a set of words
as the following equation~\ref{equ-sentence}shows:
\begin{equation} \label{equ-sentence}
S_{s\in\mathbb{R}}=\{s_1, s_2, ... ,s_n\}=\sum_{i=1}^{S}W_i
\end{equation}

The term Micro-Review
which is denoted as
MR is also defined as
a group of words denoted
as the following equation~\ref{equ-micro-review} shows:

\begin{equation} \label{equ-micro-review}
MR_{N\in\mathbb{R}}=\{W_1, W_2, ... ,W_n\}=\sum_{i=1}^{N}W_i
\end{equation}
Where N is the size
of the micro-reviews.
Moreover, the corpus
of rewires in our case
which is denoted
as $G_{_(number of reviews)}$
is defined as a collection
of reviews and micro-reviews
that are available in the
dataset which is mainly
denoted as the following
equation~\ref{equ-corpus}shows:

\begin{equation} \label{equ-corpus}
G_{C\in\mathbb{R}}=\{D_1, D_2, ... ,D_n\}=\sum_{i=1}^{C}D_i=\sum_{i=1}^{C}(\sum_{j=1}^{C_R} R_{C_R}+\sum_{k=1}^{C_{MR}}MR_{C_{MR}})
\end{equation}

Where $C_{R}$ and $C_{MR}$
are the number of
the reviews and
micro-reviews separately.
The personalization term
also refers to the fact
that the review selection
is associated with a user and
his/her preferences.
Moreover, the term
Preferences which is
denoted as P is defined
as a set of likings of user
denoted as the following
equation~\ref{equ-preferences} shows:

\begin{equation} \label{equ-preferences}
P_{P\in\mathbb{R}}=\{P_1, P_2, ... ,P_n\}=\sum_{i=1}^{P}P_i
\end{equation}

where $P_{i}$$(1 \leq i \leq n)$.

The term \textit{Preferences of
Previous Users} which is
 denoted as pp,
 is defined as set of
 preferences of previous
 users denoted as the
 following equation~\ref{equ-previous-preferences}shows.

 \begin{equation} \label{equ-previous-preferences}
PP_{PP\in\mathbb{R}}=\{PP_1, PP_2, ... ,PP_n\}=\sum_{i=1}^{P}P_i
\end{equation}

where $PP_{i}$$(1 \leq i \leq n)$.

Another term is the
Matching Function
which is denoted as F,
is defined as the following
equation~\ref{equ-matching-function} shows:

\begin{equation} \label{equ-matching-function}
F_{Score}(r,mr)=\sum_{i=1}^{R}\sum_{j}^{MR}i^{min}(r_{i}, mr_{j})
\end{equation}

where s is a sentence
in R and MR are a reviews
and micro reviews.
The function also considers
P and PP while checking similarity.
Selection Coverage which is
another term in our problem
that is denoted as
\textit{\textit{Coverge}(R)}
and it is defined as the
maximum number of
micro-reviews matching
with limited number of reviews
that satisfy user preferences
as it shown in
equation~\ref{equ-coverge}:

\begin{equation} \label{equ-coverge}
Coverge(R)=MR^{max}((|R\subseteq MR|), (\dfrac{P}{|MR|}))
\end{equation}

The main criteria of
the selection majority
in this problem is the
\textit{Selection Efficiency}
which refers to the
efficiency of a review R.
In another word,
the \textit{Selection Efficiency} is
nothing but fraction of relevant
sentences in R that satisfy user
preferences as the following equation~
\ref{equ-selection-efficiency} shows:

\begin{equation} \label{equ-selection-efficiency}
\textit{Efficiency(R)}=\dfrac{|R^{r}|}{|R|}
\end{equation}

The process of reviewing
became easier of late,
thank to web applications
and web portals where
reviewing is facilitated.
(\url{www.YELP.COM})and
(\url{www.Foursquare.com})
are best examples for reviews
and micro-reviews respectively.
With the emergence of
social networking,
the reviews and micro-reviews
on various entities are
growing exponentially.
Therefore, it is essential
to have automated mechanisms
to choose genuine and
high-quality reviews.
Such reviews can help
in making well informed
decisions or gain knowledge
quickly from the knowhow
of other people who have
already experienced
merits and demerits of
certain service rendered.
However, it is a challenging
to identify such genuine and
high-quality reviews.
Especially finding
such reviews with high
coverage is NP-hard.
Many researchers contributed
towards it.
The review selection is
studied in~\cite{GhoseI07,KimPCP06} and~\cite{lappas2010efficient}.

Recently Nguyen~{et~al.}~\cite{NguyenLT15J}
proposed a mining technique
that considers micro-reviews
and coverage of micro-reviews
on given entity as an
objective function to
discover reviews that
reflect genuine and
high coverage content
that is very useful to users.
They considered a product or
service for which reviews
are taken from Yelp.com and
micro-reviews are taken
from Foursquare.com.
They also treated
micro-review as a tip
or recommendation.
Choosing micro-reviews
related to a product as
objective function for matching
with reviews and selection
of reviews appeared meaningful
approach as the micro-reviews
of same service reflect short
recommendations.
Considering micro-reviews
as coverage made sense.
However, personalization
in review selection is
the research area
which is not yet explored.
This is the basis for
the research which needs
to investigate further to
evaluate the research
questions given below.
The main research objective
is "Personalized Review
Selection Using Micro-Reviews".
Selection of reviews
based on matching coverage
of micro-reviews needed
further optimization.
Therefore, we consider
it as an optimization problem.


\section{Proposed Methodology} \label{sec-proposed-methodology}

Social media helps to know
about user personality,
products and even services
though the customers reviews
and their interests and
emotions that can be
demonstrated as comments
during the reviewing.
In another word, reviewer's
comments can have extracted
by monitoring person's
social activities as well as
the reviewing controlling
and visualizing systems
that enable to provide
personalized data/services
and their opinion (comments).
Our proposed system aims to
facilitate the mining of
reviews and micro-reviews
in order to discover
personalized reviews
that satisfy the user preferences.
This section provides
the methodology that
used to investigate the
process of achieving personalized
review selection using micro-reviews
through our proposed system.
A framework of our
proposed system as part
of the methodology technique
which guides the research
to be carried out to
understand the whole approach.
It provides various components
that can be used to complete
the research and evaluate the work.

\subsection{Data Collector} \label{subsec-data-collector}

\begin{figure}[htbp]
\centering
  \includegraphics[width=0.9\textwidth]{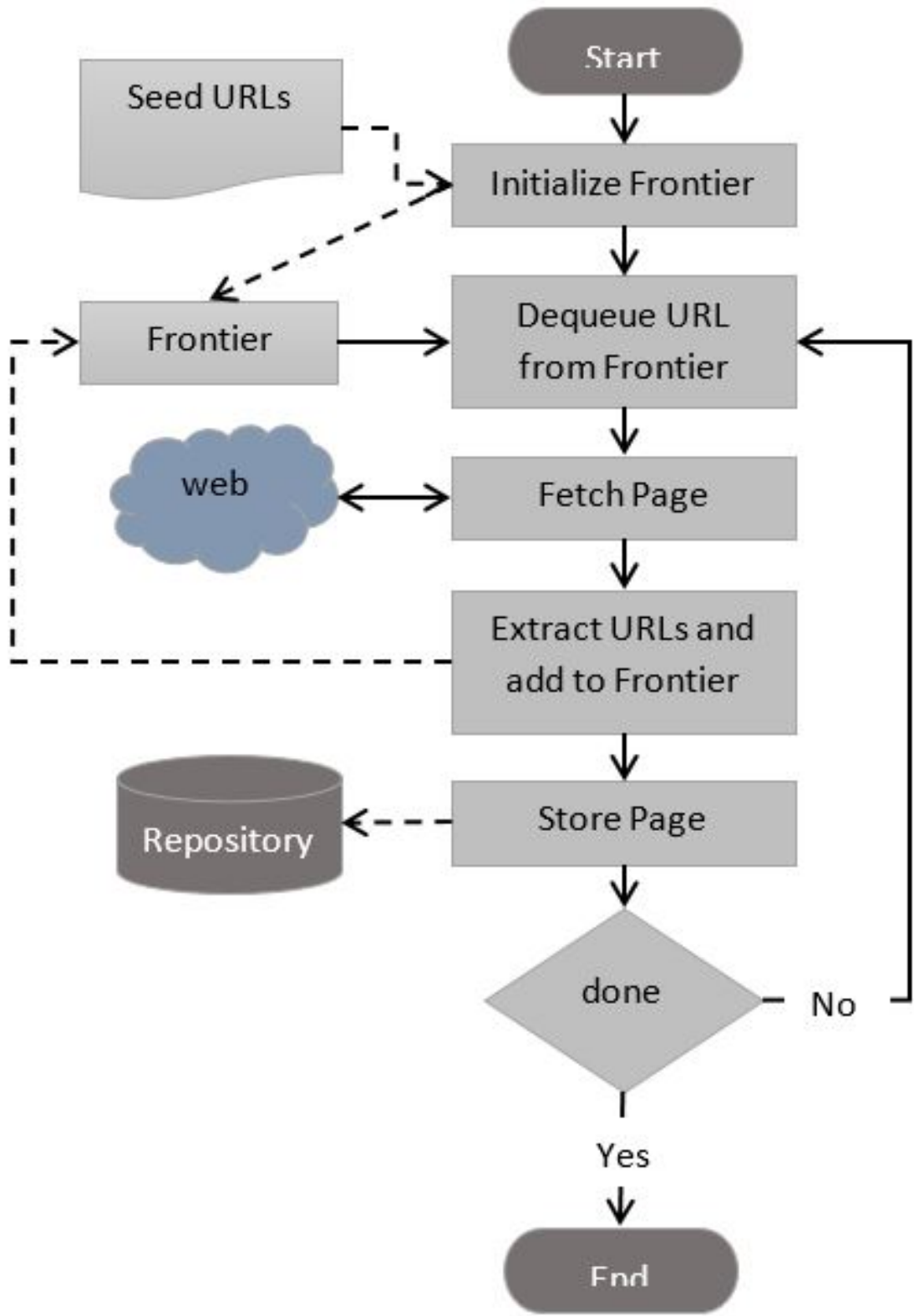}\\
  \caption{Basic crawlers Flowchart Implementation}
  \label{fig-data-collecter}
\end{figure}

The first stage of our
proposed system is the
Data collector.
The ability to find
specific information and
websites is becoming
increasingly essential
for a typical end-user
whilst browsing the web.
This is essentially one
of the main reasons as
to why crawler software
exists in which to
provide users a mirror
of the web,
either for archive purposes
or simply to find relevant
information using search keywords.
The basic model that
we used to do the data
collector is the Web
crawlers model implementation
which is illustrated in Figure~\ref{fig-data-collecter}.
In this basic model there
is a sequential web crawler
staring from the Seeds which
can be any list of starting URLs.
In order of starting page visits,
which is determined by
frontier data structure(\url{http://www.Yelp.com}).
Datasets are collected from
two sources. The first one is
the (\url{http://www.Yelp.com})
and the second
one is the (\url{http://www.Foursquare.com}).
Reviews dataset on restaurants
is collected from (\url{http://www.Yelp.com}) while
the micro-reviews
dataset which is related to the
same entities is collected from
(\url{http://www.Foursquare.com}).

\subsection{Pre-processing} \label{subsec-pre-processing}
The second stage of our
proposed system is
the pre-processing stage.
In this stage the Data
Pre-processors first
removes slang and
abbreviated words using
commonly used slang in
social media posts and Web pages.
Web pages' data also consist
of spelling mistakes which
badly effect the information
extraction process.
The pre-processing
stage in our approach
has made the datasets
per-processing to provide
appropriate data
(reviews and micro-reviews comments)
from the raw data in two phases:

\subsubsection{Stop Word Removal}
Stop words are the words
in the set of documents
(corpus) containing certain
words that do not make any
difference in the text clustering process.

\subsubsection{Stemming Process}
A stemming process which
identifies root words
and removes all derived words.
The well-known class
Porter Stemmer algorithm
is reused here for stemming mechanism.

\subsection{Profile Builder}
\begin{figure}[htbp]
  \includegraphics[width=0.9\textwidth]{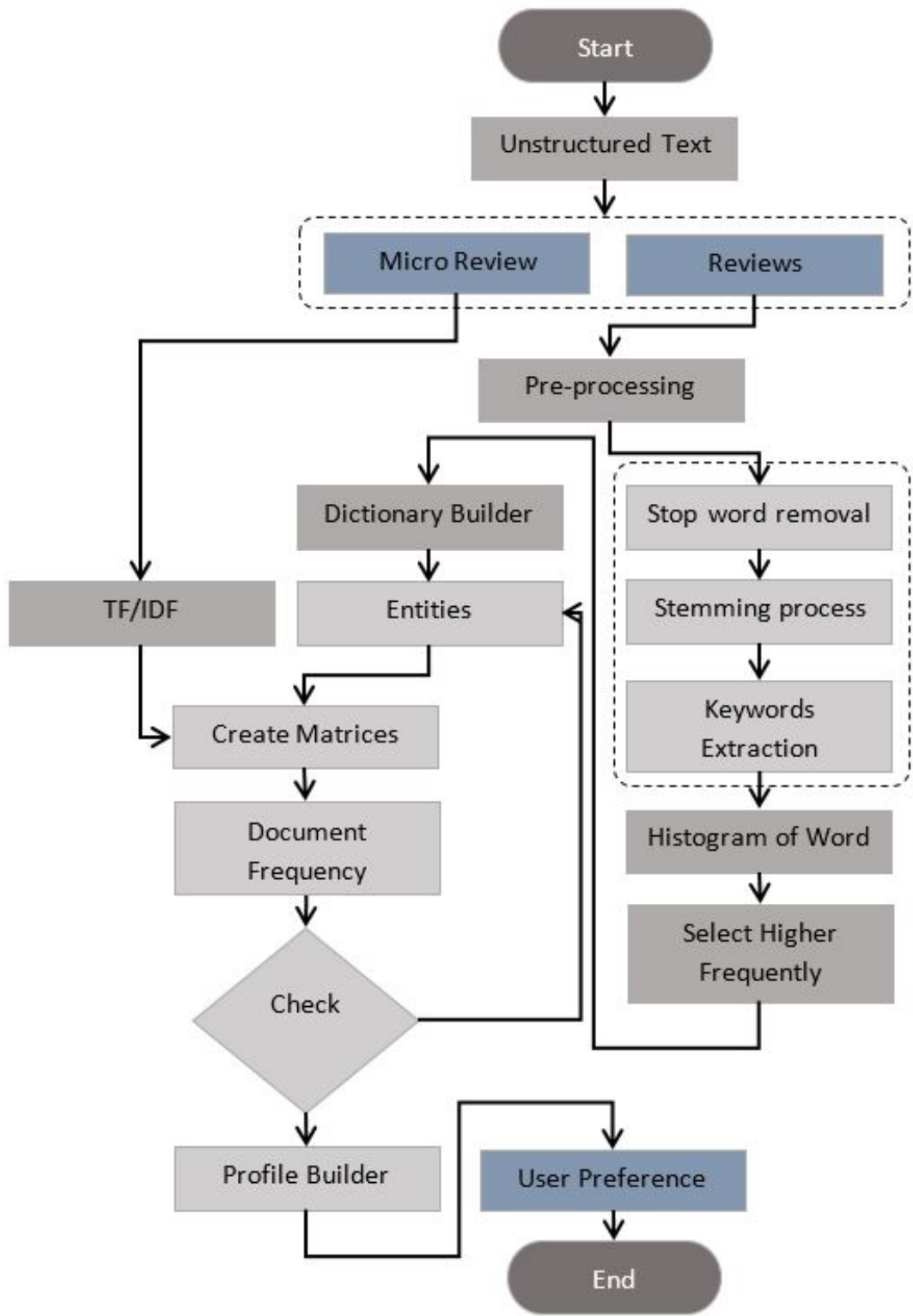}\\
  \caption{Profile Builder Flowchart Implementation}
  \label{fig-data-profile}
\end{figure}

This submodule extracts
useful information
from Web pages and
maintains history
to build user
(reviews) profile.
Our approach for the
profile builder has
three main stages.
The first stage is the
pre-processing stage
where the input for this
stage is unstructured
text data from the reviews
dataset.
In this stage the stop words
are removed which are
the words in the set
of documents (corpus)
containing certain words
that do not make any
difference in the
text clustering process.
Then, the stemming process
is done which identifies
root words and removes
all derived words.
The well-known class
Porter Stemmer algorithm
is reused here for
stemming mechanism.
The next stage is
Keywords extraction
and dictionary builder
which consists the
high frequently words
after applying the
Histogram of Words (WoF).
Moreover, Profile constructor extracts
customer's preferences by using Alchemy API.
It takes unstructured text
and acquires knowledge by presenting
the semantic richness hideaway in sentiment relevant to those entities.
The system keeps extracted user sentiments, keywords, entities
in user's profile repository for futurity utilize.
The final stage is the
profile builder by using
TF/IDF matrices creator.
In this stage, the TF/IDF
stands for
Term Frequency/Inverse Document Frequency
which is a standard measure to
reflect the importance of
a word to a document with
respect to the corpus.
Based on the return results
of the TF/IDF each
document is assigned
to one category to build
the profile for this document.
Figure \ref{fig-data-profile}
shows the main flowchart
of the profile builder that
is used in this paper.

\subsection{Sub Micro-Review Set Selection}\label{sec-user-preferences}

The user preference or
the sub micro-reviews set
selection is the main
significant step that
has been proposed in this paper.
Based on the mathematical model
of the best micro-review
set selection,
the most relevant sub-set
of micro-reviews is
going to generated and selected.
The main mathematical model
as well as the relevant
micro-review sub-set
selection are described below:

In this section,
we will explain
the necessity of
micro-review sub-set
selection in
our PeRView approach
by selecting a small set
of micro-reviews
that cover
as many reviews as possible,
with few sentences.
Mathematically,
we call the Micro-Review
$MR_{N\in \mathbb{R}}$
efficiency covers the
whole Review set
$R_{R\in \mathbb{R}}$
if a Micro-Review's sentence
 $S_{S\in MR_{S}}$
 matches the Review comment.
 In another word,
 Micro-review $MR_{N\in \mathbb{R}}$
 covers any review
 $R_{R\in \mathbb{R}}$
 if there is a sentence
 $S_{S\in MR}\in MR_{N\in \mathbb{R}}$.
 Intuitively,
 we can define the
 selection problem
 of the Micro-Reviews
 sub-set by denoting
 that the Micro-Reviews
 sub-set $T_{N\subseteq MR}$
 is a sub-set of Micro-Reviews
  $t_{MR\in \mathbb{R}}$
  that the Reviews set
   $T_{R}$ is covered
   by at least one sentence
   from the Micro-review
    $MR_{N\in \mathbb{R}}$sentences.
    Formally, we can define
    the whole problem as
    equation~\ref{equ-best}
    shows below:
\begin{equation}\label{equ-best}
T_{R}={t_{MR}\in T_{MR_{t_{MR}}}:\exists S_{S\in t_{MR}}\in MR_{N\in \mathbb{R}},R_{R\in \mathbb{R}},F(S,R)=1}
\end{equation}\\

we say that
$T_{MR}$ sub-set
covers the Review
topic $T_{R}$ by
defining the coverage
of review R as
the following
equation~\ref{equ-coverage}.
\begin{equation}\label{equ-coverage}
Cov(MR)=\dfrac{T_{MR}}{T_{R}}
\end{equation}\\

We can extend this
definition to the
case of a collection
of a subset of micro-reviews.
For a set of micro-reviews
$S\subseteq MR$,
we define the coverage
of the set $S$ as
equation~\ref{equ-cov-set}
shows below:

\begin{equation}\label{equ-cov-set}
Cov(S)=\dfrac{|\cup_{MR\in S}T_{MR} |}{T_{R}}
\end{equation}\\
\subsubsection{Efficiency Selection Criteria}\label{sub-sec-efficiency}

In some cases,
some micro-reviews
may have high coverage
which means that
many sentences that
are not relevant to
any topic of the review at all.
To avoid such similar case,
the efficiency selection
criteria should be used.
For a micro-review set
~$MR_{N\in \mathbb{R}}$
let assume that
$MR^{mr}$ of
such "relevant"
sentences which cover
at least one review topic
as show in~\ref{equ-eff}:
\begin{equation}\label{equ-eff}
Mr^{mr}={S_{s\in MR}\in MR_{N\in \mathbb{R}}:\exists t_{MR}\in T_{MR}, F(S,t)=1}
\end{equation}

Then, the define the
efficiency \textit{efficiency(MR)}
is define as a fraction of
"relevant" sentences in MR.
Which is formally can
be written as the
equation~\ref{equ-eff-mr} shows:
\begin{equation}\label{equ-eff-mr}
\textit{Eff}(MR)={\dfrac{|MR^{mr}|}{|MR|}}
\end{equation}

Extending the definition
of efficiency to a
collection of micro-reviews
is a little more involved.
We need a way to
aggregate the efficiency
of the individual
micro-reviews.
In this case
we use the average efficiency
of a set S is defined
as the average efficiency
of the micro-reviews
in the set.
Formally, we can define
that as the following
equation~\ref{equ-eff-average}:
\begin{equation}\label{equ-eff-average}
Eff_{Average(S)}={\dfrac{\sum_{MR\in S}Eff(S)}{|S|}}
\end{equation}\\

The algorithm,
shown in Algorithm~\ref{alg-micro},
proceeds in iterations
each time adding one
review to the collection \textit{S}.
At each iteration,
for each review \textit{MR}
we compute two quantities.
The first is the gain gain(\textit{MR}),
which is the increase
in coverage that we obtain
by adding this micro-review
to the existing collection \textit{S}.
The second quantity
is the cost Cost(\textit{R})
of the review \textit{MR},
which is proportional to the
inefficiency $1-\textit{Eff(R)}$
of the review,
that is,
the fraction of sentences
of \textit{MR} that
are not matched
to any review.
We select the micro-review
$MR^{*}$ that has the
highest gain-to cost
ratio and guarantees
that the efficiency of
the resulting collection
is at least $\alpha$,
where $\alpha$ is a parameter
provided in the input.
The intuition is that
reviews with high
gain-to-cost ratio cover
many additional tips,
while introducing little
irrelevant content,
and thus they should
be added to the collection.

\subsubsection{\textit{Micro-Reviews Selection}} \label{sub-sec-micro-algorithm}

\begin{algorithm}
\scriptsize
\caption{Micro-Reviews Selection Algorithm}
\label{alg-micro}
\begin{algorithmic}[1]
\REQUIRE
      Set of Micro-reviews MR,
      Set of Reviews R,
      Efficiency function Eff;
            Threshold T:
            selection number of the micro-reviews,
            parameters $\alpha,\beta$

\ENSURE
      Set of Micro-reviews $ S\subseteq MR$ of size T.
\STATE S=0;
\IF{$ |S|<T$}
\FOR {all $MR\in R$}
\STATE gain (MR)= Cov($ S \cup MR$)-Cov(S)
\STATE Cost(MR)= $ \beta (1- Eff(MR))+(1-\beta)$
\ENDFOR
\STATE $ \varepsilon ={MR \in R: Eff(S \cup MR)\geq 0}$
\IF {$ (\varepsilon ==0) or~max_{MR\in _{\varepsilon}}gain(MR)==0$}
\STATE Break
\ENDIF
\STATE $MR^{*}=ar gmax_{MR\in _{\varepsilon}} gain(MR)/cost(MR)$
\STATE $ S=S\cup MR$
\STATE $ R=R/MR^{*}$
\ENDIF
\RETURN S
\end{algorithmic}
\end{algorithm}

\begin{figure}[htbp]
  \includegraphics[width=0.9\textwidth]{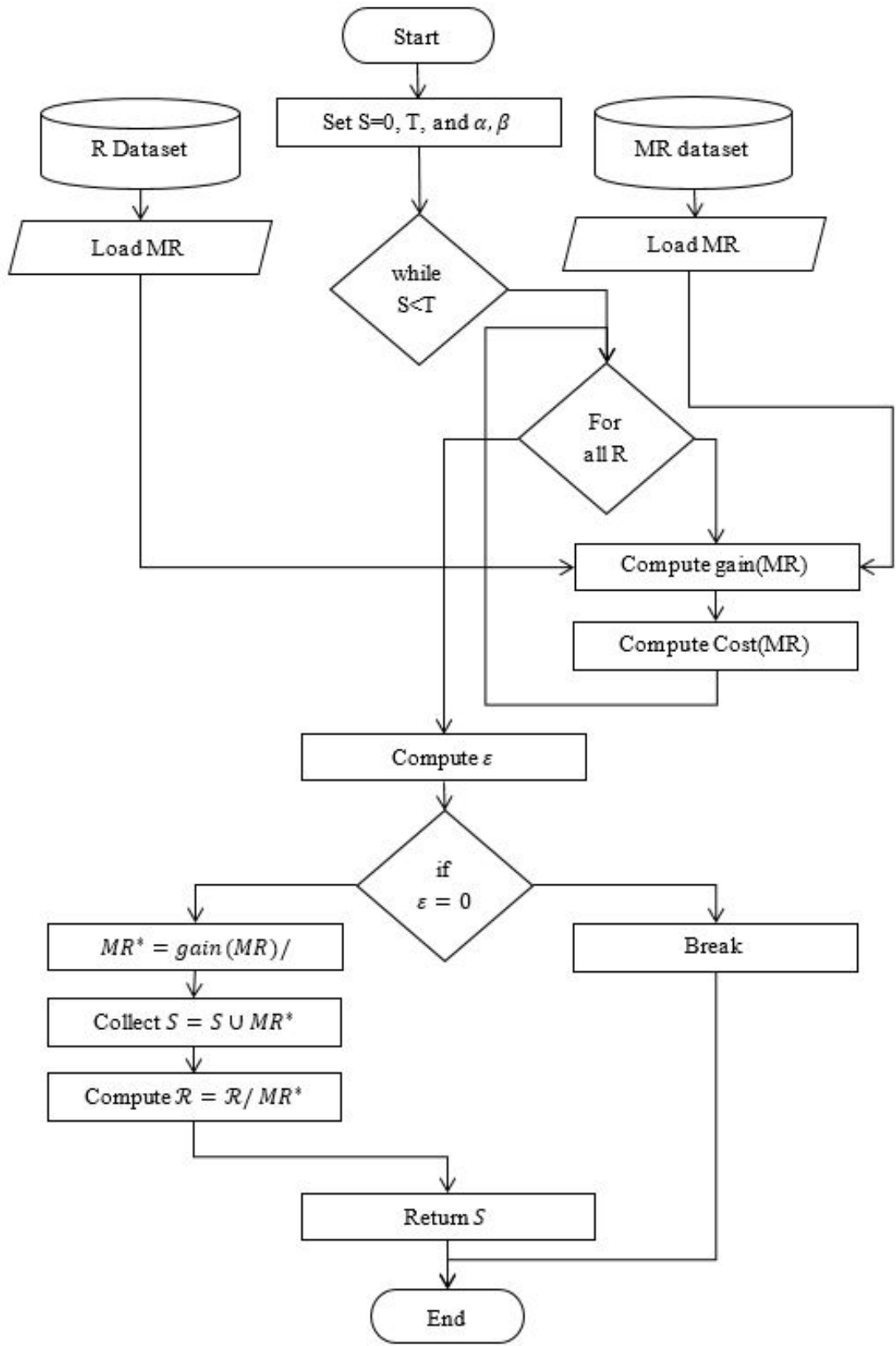}\\
  \caption{Micro-reviews sub-set selection}
  \label{fig-micro-algo}
\end{figure}

The whole system
flowchart of the
micro-reviews selection
is illustrated in
the Figure~\ref{fig-micro-algo}.

\subsection{Our Proposed Framework Personalized Reviews Selector  (PeRView)}

\begin{figure}[htbp]
  \includegraphics[width=0.9\textwidth]{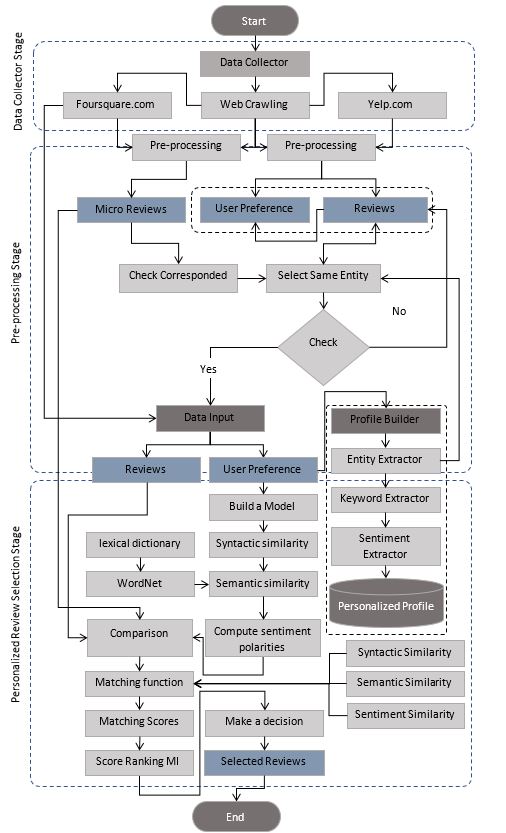}\\
  \caption{Overview of PeRView framework}
  \label{fig-system}
\end{figure}

The proposed framework that
we claimed in this paper is
called PeRView which is mainly
designed to facilitate
the mining
of reviews and micro-reviews
in order to discover
personalized reviews
that satisfy the user
preferences
is illustrated in
Figure\ref{fig-system}.
Our proposed system
framework has three
main stages.
The first stage is
the data collector stage
which is used to collect
all the reviewers,
micro-reviewers,
while the second stage
is the pre-processing stage.
Finally, the third stage
of our proposed system
is the Personalized Review
Selection Algorithm (PRSA)
which captures user preferences
and model them although
it performs comparison
of micro-reviews with
reviews sentence and
compute the syntactic similarity score.
Then, the final decision
of Reviewer selection
will make based on
using the Mutual Information
Ranking Mechanism (MIM)
which is used to rank
the similarity scorning
to select the best one.

The proposed framework
takes input as micro reviews,
and their corresponding
that related to same entity.
Also, it takes the reviews
and user preferences
as another input in
the whole input stage
which is done by doing
the data collector stage.
Then, it makes use of
the proposed algorithm named
\textit{Personalized Review Selection Algorithm (PRSA)}
which performs various
activities on the given inputs.
First of all,
it captures user preferences
and model them.
Afterwards it performs comparison
of micro-reviews with reviews
sentence wise by employing
syntactic similarity.
Then it employs semantic
similarity using lexical
dictionary known as
WordNet~\cite{manning1999foundations}.
The topic modelling concept is
employed by using \textit{Mutual Information
Ranking mechanism (MI)}
which is used to rank
the similarity scorning
to select the best one.
Then sentiment polarities
of micro reviews and
reviews are compared
to make a decision
to include them.
Afterwards, the matching function
is applied in order to
discover personalized reviews
that reflect high coverage,
high quality and satisfy
user preferences.



\begin{algorithm}
\scriptsize
\caption{Personalized Review Selection Algorithm}
\label{alg-PRSA}
\begin{algorithmic}[1]
\REQUIRE
       Set of reviews \textbf{R},
       set of micro-review \textbf{MR}
       related to \textbf{R,}
       user preferences \textbf{P,}
       threshold \textbf{t}
\ENSURE
      Set of reviews that with high quality and coverage \textbf{R'}
\STATE initialize \textbf{R'} to hold result
\FOR {each review \textbf{r} in \textbf{R}}
\FOR {each sentence \textbf{s} in \textbf{r}}
\STATE $count\leftarrow0$
\FOR {each micro review \textbf{mr} in \textbf{MR}}
\STATE $sim_1\leftarrow find-Syntactic-Sim(s, mr)$
\STATE $sim_2 \leftarrow find-Semantic-Sim(s, mr)$
\STATE $sim_3 \leftarrow find-Sentiment(s,mr)$
\STATE $sim\leftarrow Rankink-evaluation(sim_1, sim_2, sim_3)$
\STATE $sim\leftarrow Selection(sim_1, sim_2, sim_3)$
\IF {\textbf{s} covers \textbf{mr} and \textbf{P} with \textbf{sim}}
\STATE increment $\rightarrow count$
\ENDIF
\ENDFOR
\ENDFOR
\IF {$count \geq t$}
\STATE add review $\rightarrow \textbf{R'}$
\ENDIF
\ENDFOR
\RETURN R'
\end{algorithmic}
\end{algorithm}

Algorithm~\ref{alg-PRSA} takes set of reviews
of chosen restaurant,
set of micro-reviews of same restaurant,
user preferences for personalization
and matching threshold as input and
produces set of reviews that
exhibited high quality and coverage.
For each review all micro-reviews are
compared for similarity.
Similarity is found syntactically,
semantically and polarities related to sentiments.
Then the similarities are merged
to have final quantitative value.
The number of micro-reviews matching
with review as much as possible determines its coverage.
This is verified by suing given threshold
and the reviews are finally selected.
Syntactic similarity
is achieved by using keyword
similarity model found in~\cite{manning1999foundations}.
Each sentence in a review
denoted as \textit{\textbf{s}}
and each micro-review
denoted as \textit{\textbf{mr}}
are associated with a vector
as it shown in equation~\ref{equ-syntactic}:\\

\begin{equation}\label{equ-syntactic}
Syntactic Similarity(s,mr)=\cos(s,mr)
\end{equation}\\

Since syntactic similarity
is based on keywords present in
the review and micro-review,
standard $TF-IDF$ (Term Frequency – Inverse Document Frequency)
approach is used to know relative importance of a word.
With respect to semantic similarity
a review and micro-review are similar
if they describe same concept though
the words are different.

Associates topic of each micro-review
with probability distribution of topics.
It shows which are very important for given micro-review.
The similarity of topic distributions
is used as measure semantic similarity between
a micro-review and a review sentence.
The distance measure between two probability distributions
used is Jensen-Shannon Divergence.
If divergence is more,
it indicates lesser in similarity
as it shown in equation~\ref{equ-semantic}:\\

\begin{equation}\label{equ-semantic}
Semantic Similarity(s,mr)= 1-JSD(\Theta_s,\Theta_mr)
\end{equation}\\

Where\textbf{ s} refers to sentence in
a review and \textbf{mr} refers to a micro review.
Probability distribution of sentence and
micro review are denoted as $\theta$s and $\theta$mr respectively.
Then the third similarity is sentiment similarity.
It refers to the fact that sentiment
of review and micro-review should match.
Sentiment polarities are computed as
discussed in~\cite{chen2017modeling}
and shown in equation~\ref{equ-sentiment}:\\

\begin{equation}\label{equ-sentiment}
Sentiment Similarity(s,mr)= Polarity(s)\times Polarity(mr)
\end{equation}\\

By merging these three similarity measures,
a final matching function is
computed which returns either $0$ or $1$
indicating binary decisions
such as not matching and matching.
The similarity knowhow is used
to have a binary classifier which
determines match or not match
based on the given threshold.\\

\section{Experimental Results}\label{sec-exp-results}
The experimental results of the proposed system are based mainly on the objective experiments which shows the effeteness of the proposed approach. Finding a set of micro-reviews that covers as many reviews as possible which is used later to select the best reviews is the main contribution of the proposed system.  In this section, first, we describe the datasets that are used in the experimental results. This is followed by the evaluation metric that is used to select the best set of reviews. Then, we discuss the experimental results that the proposed system is achieved.

\subsection{Dataset}\label{subsec-dataset}
The reviews and micro reviews dataset have been collected from different websites such as (\url{www.YELP.COM})and
(\url{www.Foursquare.com}) which are used for experiments. Observations are made with matching percentages of reviews with each micro review.  Dataset corpus is collected from two sources. They are (\url{www.YELP.COM}) and (\url{www.Foursquare.com}). Reviews on restaurants are collected from~(\url{www.YELP.COM}) while the micro-reviews related to the same entities are collected from~(\url{www.Foursquare.com}). Several datasets pertaining to different restaurants for which reviews and micro-reviews available are collected from those sources.\\

\subsection{Evaluation Metric}\label{subsec-evaluation-matric}

In case of evaluate select the most significant set of reviews that has the highest Personalized coverage score, there are two main factors that we have include in our elevation and selection formula. Personalized matching score and reviews set size are the main factor that our evaluation matric should considers. In such a similar case, our evaluation problem definition differs depending on the choice the minim set that has the higher personalized coverage score. In this case, we define the efficiency personalized scoring function ${\textit{Per-Eff}_{min}}$ which is defined as the minimum set of selected reviews \textbf{S} that has a higher personalized coverage score. Formally, we define the minimum efficiency selected set by:

\begin{equation}\label{equ-per-eff}
{\textit{PerEff}_{min}}=min_{R\in S}(R)
\end{equation}

The $Personalized-Eff_{min}$ in our evaluation case problem is that each individual Personal selected review must have a personalized similarity score that by computing the personalized efficiency score should be at least $\alpha$ as a constraint which presents the personalized coverage evaluation score.  Formally we define the that as:

\begin{equation}\label{equ-personalized-eff}
{\textit{PerEff}_{min}}(S)\geq \alpha
\end{equation}

Since the size of the personalized selected reviews set is the second factor that our evaluation should consider it. Therefore, we define such an optimization function called maximization of the personalized coverage score  \textit{MaxPerCoverage}, where the reviews personalized similarity scores are restricted to the personal selected review's subset size in such should have an efficiency personalized score at least $\alpha$. In this case the \textit{MaxPerCoverage} optimization function can be used for obtaining an optimal evaluation solution.

In case of define the \textit{MaxPerCoverage} optimization evaluation function, let assume that $X_{i}$ is obtained as personalized similarity score that associated with each personal review $R_{i}$ and based on that each individual personal review has been selected. Also, denoting that $R_{i}$  is the personalized selected set. Although, let's assume that $Y_{i}$ is the sub personalized review size that associate with each personal selected set $S_{i}$. We also define another parameter called $C_{i}$ that associated with each personal subset $S_{i}$, with $C_{i}$=1 denoting that selected such a personal subset $S_{i}$ is covered by at least one of review (personalized matching score) in the selected set, and $C_{i}$=0 in the otherwise.

Intuitively, to maximize the \textit{MaxPerCoverage} score for any personalized selected subset based on the main factors that we have defined above, we define the optimization problem in our case as a problem with a set of constraints to maximize the \textit{MaxPerCoverage} evaluation score such as:

\begin{equation}\label{equ-maxper}
maximize\sum_{j=1}^{m} C_{i}=Per_Simalirity(s, mr)
\end{equation}

The evaluation objective function as it is defined in~\ref{equ-maxper} maximizes the \textit{MaxPerCoverage} evaluation score based on some constraint. The first constraint is defined in ~\ref{equ-supject} which ensues the number of selected reviews (sub-personalized set size) does not exceed $\textit{K}$ which means at least one personal review covers the whole set.

\begin{equation}\label{equ-supject}
subject~ to\sum_{i=1}^{n} X_{i}\leq K
\end{equation}

In additional to that, the second constraint is defined in~\ref{equ-size} which ensures that average personalized similarity matching scores (that covers the whole selected set $S_{i}$ based on the personalized average score) and also based on the size of its set $Y_{i}$ to compute the average personalized similarity scoring at least one review that covers $C_{i}$ must be selected.

\begin{equation}\label{equ-size}
\sum_{i:S_{j}\in S, R_{i}}^{n} {\dfrac{\sum_{i=1}^{n}X_{i}}{Y_{i}}} \geq C_{i}\forall s_{j}\in T
\end{equation}

 Other constraints which are define in~\ref{equ-x} and~\ref{equ-c} are required to ensure that the value of both average personalized similarity score as well as the \textit{MaxPerCoverage} evaluation score between the range 0 and 1:

 \begin{equation}\label{equ-x}
X_{i}=\{0,1\}
\end{equation}

\begin{equation}\label{equ-c}
C_{i}=\{0,1\}
\end{equation}

Therefore, it is well known that the main aims of the \textit{MaxPerCoverage} optimization function is to always select the personal review set whose adding maximizes the personalized coverage score and it should be closer than the other reviews based on the approximation personalized ratio for the \textit{MaxPerCoverage} which is formally defined in the following equation:

\begin{equation}\label{equ-approximation}
\textit{Approximation Personalized_ratio}=(1-{\dfrac{1}{\ell}})
\end{equation}

Where $\ell$ is defined as the natural logarithm.  In other word, we define the \textit{MaxPerCoverage} score is based technically on the $\textit{Per-Eff}_{min}$ efficiency score that has approximation personalized ratio as it defied in~\ref{equ-approximation}. Finally, we present the \textit{MaxPerCoverage} algorithm that basically based on many iterations that in each iteration by adding new review (based on the personalized similarity score) on the collection set $S_{i}$ based on the personalized matching $X_{per_{i}}$ score and the set size $S_{i}$  that is automatically updated upon each new review has been added to the set.  In this case, to evaluate each subset, in each iteration that we select one individual review $\textbf{R}$ based on the personalized matching score between the closest reviews $R_{R\in \mathbb{R}}$ set and the micro-reviews set $MR_{MR\in \mathbb{R}}$ we compute two quantity scores. The first score is the information gain based of the personalized selected review $\textit{gain}(R_{per})$ which increase the maximization the personalized coverage score based on the average personalized similarity score and the size of the reviews set.  Formulary, it is defined in the~\ref{equ-gain}.

\begin{equation}\label{equ-gain}
PerGain(R_{set})=(\dfrac{\sum_{i=1}^{n=size(y_{i})}X_{per_{i}}}{n})
\end{equation}

Which in our maximization optimization case~\ref{equ-cost} will increase the \textit{MaxperCoverage}  score by adding each new closest review (personalized one) to the collected set $S_{i}$ based on its personalized similarity score.

The second quantity score is the $Cost(R_{per})$ of the personal review $\textbf{R}$ which is mathematically presents the efficiency personalized score that based on the next formula.

\begin{equation}\label{equ-cost}
PerCost(R_{set})=1-(PerEff_{min}=min_{R_{set}\in S}(R_{set}))
\end{equation}

Finally, we select the personalized review set that get the highest gain and cost ration based on the personalized matching similarity score as well as the size of the personalized reviews set. The \textit{MaxPerCoverage} algorithm that is mainly designed to evaluate and select the best personalized review result set is described in algorithm~\ref{alg-maxpercoverage} below.

\begin{algorithm}
\scriptsize
\caption{\textit{MaxPerCoverage} Evaluation Algorithm}
\label{alg-maxpercoverage}
\begin{algorithmic}[1]
\REQUIRE
      $R_{i}$ selected review,
      $x_{i}$ personalized matching score,
      $y_{i}$ initial reviews set size;

\ENSURE
      Selected Personal Reviews set $\textit{PerC}_{S}$ that has the highest personalized evaluation score $\textit{PerC}_{R_{set}}$.
\STATE Set the total number of Selected set $\textit{T}={\sum_{i=1}^{n} S_{i}}$
\STATE Set the set size for each Reviews set $\textit{Y}_{i}=|S_{i}|$
\FOR {all $S_{i}$}
\STATE Compute the Personalized Evaluation Scores for each set based on
\STATE $PerC_{R_{set}}=\{gain(R_{set}), Cost(R_{set})\}$
\STATE Compute the Personalized Gain score for each set based on
\STATE $PerGain(R_{set})=(\dfrac{\sum_{i=1}^{n=size(y_{i})}X_{per_{i}}}{n})$
\STATE Compute the personalized Cost score for each set based on
\STATE $ PerCost(R_{set})=1-(PerEff_{min}=min_{R_{set}\in S}(R_{set}))$
\ENDFOR
\STATE Get the size of the PerGain$(R_{set}) scores C_{R_{set}}$
\STATE Get the size of the PerCost$(R_{set}) scores C_{R_{set}}$
\STATE Evaluate the set of Pergain $(R_{set}) and Cost(R_{set})$
\FOR {$K1=1 to all PerGain(R_{set})$}
\FOR {$K2=1 to all PerCost(R_{set})$}
\IF {$Pergain(R_{set}) and PerCost(R_{set})~is ~the ~personalized ~highest ~scores,$}
\STATE Get the index of the Personalized Reviews set $PerC_{R_{set}}=index(R_{set})$
\ENDIF
\ENDFOR
\ENDFOR
\RETURN $PerC_{R_{set}}$
\end{algorithmic}
\end{algorithm}

\subsection{Matching}\label{subsec-matching}
In our experimental results, the personalized matching between any review and the micro-review is the main objective of the proposed system. In this case, the proposed system is mainly aiming to established that by achieving a reasonable level of quality personalized matching which the proposed system is represented by the \textit{threshold personalized coverage} value. The review in this case is selected based on the personalized coverage selection algorithm that we have discuss which would be a good reflection of main review personalized coverage scope.\\

\begin{figure}[htbp]
  \includegraphics[width=0.9\textwidth]{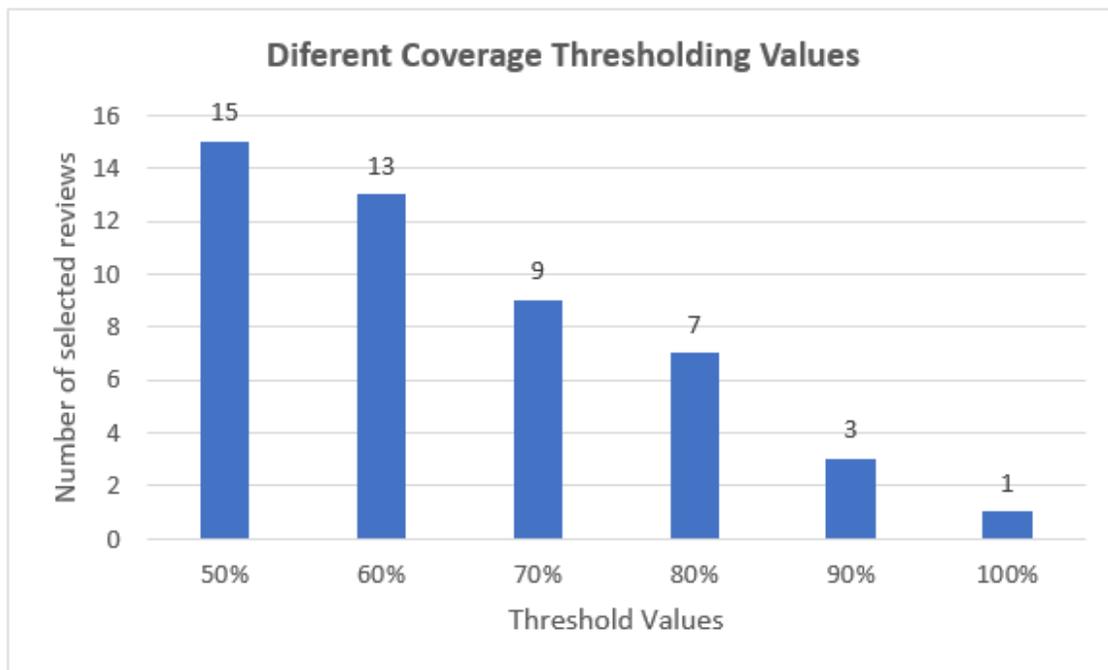}\\
  \caption{Personalized Matching (Coverage Thresholds) Results}
  \label{fig-covthreshold}
\end{figure}

In our experimental result we choice different coverage threshold score in case of showing the probability of the personalized matching. Figure~\ref{fig-covthreshold}shows the different threshold value selection based on the personalized review's coverage scoring (personalized matching score). We can see that as much as our threshold value selection is significant strong personalized sub-set such as $(90\%)$ or $(100\%)$ the number of the personalized reviews selection will monotonically decreased. As it is shown in Figure~\ref{fig-covthreshold}, the number of the personalized reviews' selection number is decreased from 15 (reviews) when the threshold value is $50\%$ (personalized coverage-matching score) to 13 (reviews) when the threshold value is increased to be $60\%$.

\begin{figure}[htbp]
  \includegraphics[width=0.9\textwidth]{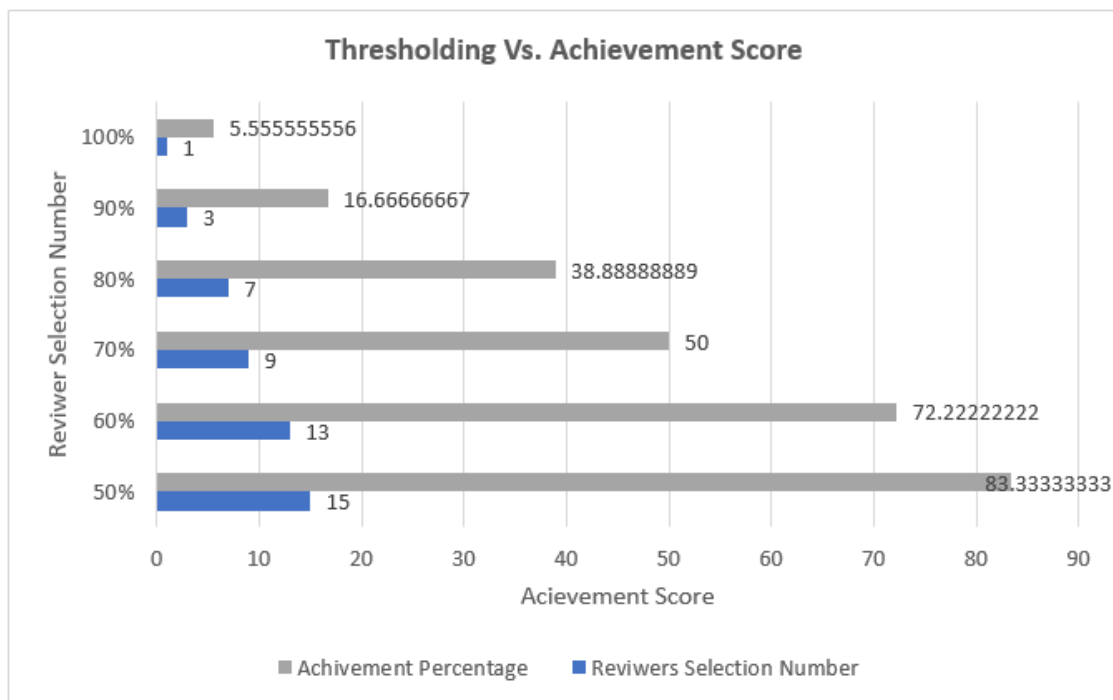}\\
  \caption{Personalized Matching Score and the Percentage of the Achievement Accuracy}
  \label{fig-achiv}
\end{figure}

It is significantly clear that the coverage thresholds value is affect the percentage of the achievement score (personalized selection accuracy) as it is shown in Figure~\ref{fig-achiv} where the best accuracy score for the best personalized reviews selection was in the threshold value (50\%) where the total number of the personalized selected reviews is 15 out of dataset reviews which achieve approximately 83.33\% accuracy. Then, the achievement accuracy is being decreased among all the other threshold values to reach to (5.55\%) in the threshold value (100\%) which is significantly tough, but our proposal system is still able to select the best personalized reviews-based on that criteria.

\subsection{Coverage and Personalized Efficiency Selection}\label{subsec-covpereff}

In term to clarify the investigation stage of the proposed system for the personalized review selection. In this section, we describe the effeteness of the \textit{MaxPerCoverage} Evaluation Algorithm on selection the best personalized reviews based on both the personalized average matching score and the reviews set size. Table~\ref{my-label} shows an example on the MaxPerCoverage scoring algorithm based on select a threshold value (90\%). During the experimental results, we can notice that the personalized reviewers set that has been select has satisfied about (61.74\%) as an average \textit{MaxPerCoverage} score based on selecting (90\%) \textit{MaxPerCoverage} threshold value.

\begin{table}[]
\centering
\small
\caption{Example Reviews Selection based on the \textit{MaxPerCoverage} Scoring Algorithm}
\label{my-label}
\begin{tabular}{cccc}
\hline
\textbf{ID}              & \textbf{Review Selection Sentence}                                                                                                                                                                                                                                                                                                                                                                                                                                                                                                                                                                                                                                                                                                                                                                                                                                                          & \textbf{Threshold}        & \textbf{MaxPerCoverage}    \\ \hline
\multicolumn{1}{|c|}{1}  & \multicolumn{1}{l|}{\begin{tabular}[c]{@{}l@{}}"Me and my friends stayed there this\\ weekend and it was GREAT. \\ Like everyone says, from the outside it looks\\ outdated (just like you see in the pictures),\\  but the rooms are in great shape.\\ They are clean and comfortable. \\ 4 of us stayed in one room and we had plenty of\\ space. The bathroom was clean and\\  they provide the needed amenities.\\ ,Also, the owners are amazing! They gave us\\ rides back and forth to the festival \\ and we never had to wait long (the motel\\ is about 10 minutes away, \\ a little more with the Coachella traffic). \\ We alsoreally enjoyed talking to them on the rides, \\ they are very fun and nice.,It is a bit \\ pricey, but so is every other\\ hotel/motel on Coachella weekends....etc.\end{tabular}}                                                                & \multicolumn{1}{c|}{90\%} & \multicolumn{1}{c|}{0.909} \\ \hline
\multicolumn{1}{|c|}{2}  & \multicolumn{1}{c|}{Not Selected}                                                                                                                                                                                                                                                                                                                                                                                                                                                                                                                                                                                                                                                                                                                                                                                                                                                           & \multicolumn{1}{c|}{}     & \multicolumn{1}{c|}{0.5}   \\ \hline
\multicolumn{1}{|c|}{3}  & \multicolumn{1}{c|}{Not selected}                                                                                                                                                                                                                                                                                                                                                                                                                                                                                                                                                                                                                                                                                                                                                                                                                                                           & \multicolumn{1}{c|}{}     & \multicolumn{1}{c|}{0.388} \\ \hline
\multicolumn{1}{|c|}{4}  & \multicolumn{1}{c|}{Not selected}                                                                                                                                                                                                                                                                                                                                                                                                                                                                                                                                                                                                                                                                                                                                                                                                                                                           & \multicolumn{1}{c|}{}     & \multicolumn{1}{c|}{0.625} \\ \hline
\multicolumn{1}{|c|}{5}  & \multicolumn{1}{c|}{Not selected}                                                                                                                                                                                                                                                                                                                                                                                                                                                                                                                                                                                                                                                                                                                                                                                                                                                           & \multicolumn{1}{c|}{}     & \multicolumn{1}{c|}{0.857} \\ \hline
\multicolumn{1}{|c|}{6}  & \multicolumn{1}{c|}{Not selected}                                                                                                                                                                                                                                                                                                                                                                                                                                                                                                                                                                                                                                                                                                                                                                                                                                                           & \multicolumn{1}{c|}{}     & \multicolumn{1}{c|}{0.727} \\ \hline
\multicolumn{1}{|c|}{7}  & \multicolumn{1}{c|}{Not selected}                                                                                                                                                                                                                                                                                                                                                                                                                                                                                                                                                                                                                                                                                                                                                                                                                                                           & \multicolumn{1}{c|}{}     & \multicolumn{1}{c|}{0.8}   \\ \hline
\multicolumn{1}{|c|}{8}  & \multicolumn{1}{c|}{Not selected}                                                                                                                                                                                                                                                                                                                                                                                                                                                                                                                                                                                                                                                                                                                                                                                                                                                           & \multicolumn{1}{c|}{}     & \multicolumn{1}{c|}{0.611} \\ \hline
\multicolumn{1}{|c|}{9}  & \multicolumn{1}{c|}{Not selected}                                                                                                                                                                                                                                                                                                                                                                                                                                                                                                                                                                                                                                                                                                                                                                                                                                                           & \multicolumn{1}{c|}{}     & \multicolumn{1}{c|}{0.8}   \\ \hline
\multicolumn{1}{|c|}{10} & \multicolumn{1}{c|}{Not selected}                                                                                                                                                                                                                                                                                                                                                                                                                                                                                                                                                                                                                                                                                                                                                                                                                                                           & \multicolumn{1}{c|}{}     & \multicolumn{1}{c|}{0.625} \\ \hline
\multicolumn{1}{|c|}{11} & \multicolumn{1}{l|}{\begin{tabular}[c]{@{}l@{}}"City Center Motel is the BEST place to\\ stay if you are in town. I recently stayed here \\ for Coachella 2014, and all I\\ can say is WOW. The rooms were amazing,\\  with a more updated look and nice TV.\\ In addition to the nice room, the service \\ i received was more than i could ask\\ for. The owners were very nice and\\  tried their very best to make our experience\\ here memorable. During the festival \\ they even offered us a ride to and from the\\ festival whenever we needed it which \\ was awesome because it was only 10 minutes\\ away and they knew how to avoid all \\ the main traffic. Overall, i would\\ recommend staying here when in town, \\ it surely was worth it. Although the\\ outside of the motel looked outdated, \\ the saying goes, ""never judge\\ a book by its covers""\end{tabular}} & \multicolumn{1}{c|}{}     & \multicolumn{1}{c|}{1.0}   \\ \hline
\multicolumn{1}{|c|}{12} & \multicolumn{1}{c|}{Not selected}                                                                                                                                                                                                                                                                                                                                                                                                                                                                                                                                                                                                                                                                                                                                                                                                                                                           & \multicolumn{1}{c|}{}     & \multicolumn{1}{c|}{0.4}   \\ \hline
\multicolumn{1}{|c|}{13} & \multicolumn{1}{c|}{Not selected}                                                                                                                                                                                                                                                                                                                                                                                                                                                                                                                                                                                                                                                                                                                                                                                                                                                           & \multicolumn{1}{c|}{}     & \multicolumn{1}{c|}{0.625} \\ \hline
\multicolumn{1}{|c|}{14} & \multicolumn{1}{c|}{Not selected}                                                                                                                                                                                                                                                                                                                                                                                                                                                                                                                                                                                                                                                                                                                                                                                                                                                           & \multicolumn{1}{c|}{}     & \multicolumn{1}{c|}{0.875} \\ \hline
\multicolumn{1}{|c|}{15} & \multicolumn{1}{l|}{\begin{tabular}[c]{@{}l@{}}"I was driving from Sacramento to Phoenix\\ and decided to stop overnight in Indio (no reservations). \\ I drove around late\\ at night to find a place to sleep \\ and learned there was a big dog show in town\\ and many motels had no vacancies.\\  Luckily, I stopped at the Bhakta's City\\ Center Motel. It's not instantly available \\ from the freeway, but it's a little\\ dream place. My room was updated\\  in every way: bed, linens, bathroom fixtures,\\ flooring, and more. There was \\ a nice/large flat screen TV on the wall, nice toiletries\\ in the bathroom, and a microwave\\  and refrigerator. This is ""the\\ season"" in Indio and my room\\  was \$69.,The family/owners are \\ extremely friendly and\\ helpful...and available 24/7....etc.\end{tabular}}                                                  & \multicolumn{1}{c|}{}     & \multicolumn{1}{c|}{0.909} \\ \hline
                         & \multicolumn{2}{c}{\textit{\textbf{Average MaxperCoverage Score}}}                                                                                                                                                                                                                                                                                                                                                                                                                                                                                                                                                                                                                                                                                                                                                                                                                                                      & \textbf{61.74\%}           \\ \hline
\end{tabular}
\end{table}

\begin{figure}[htbp]
  \includegraphics[width=0.9\textwidth]{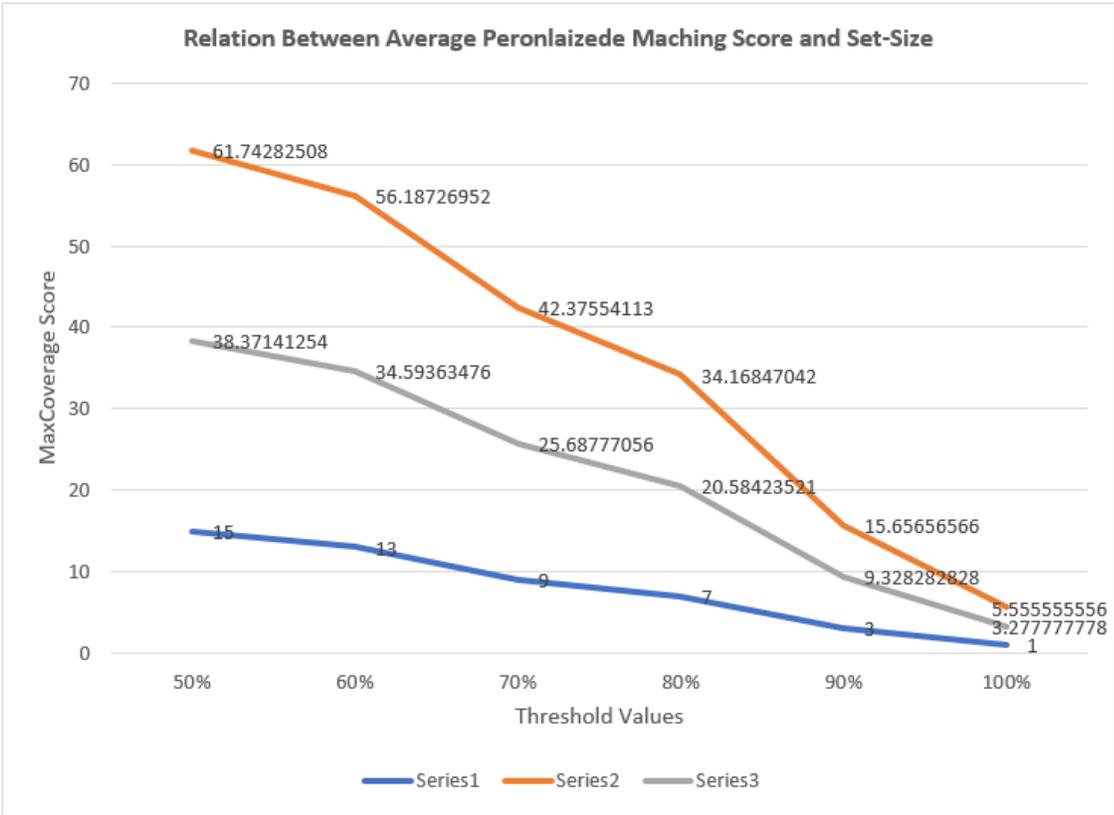}\\
  \caption{MaxPerCoverage based on the Average Personalized Matching Score and Reviewers Set-size}
  \label{fig-relation}
\end{figure}

Comparing with the other thresholds values, Figure~\ref{fig-relation} shows the relation between the set size and the average personalized matching score as it has been base on through the MaxPerCoverage selection algorithm.

We can notice that as much as the average personalized matching score for the whole sub-set is highly coverage and the personalized reviews set size is used to evaluate our selection accuracy, the performance results will be monotonically increased.  As it is shown in Figure ~\ref{fig-relation} the highest MaxperCoverage accuracy was (61.74\%) based on the total average personalized matching score and set size which is (15 reviews). Based on the set-size, two optimal and best personalized review's sets will be evaluated. One with the size 15 which achieves 61.74\% while the other one with size 13 which achieves 56.18\%.\\

\section{Conclusion} \label{sec-xyz}
In the wake of proliferation of review contents over Internet
and their importance in the contemporary world
for decision making it is essential to
choose high quality personalized reviews that are mainly consist on the main reviewed topic.
In this paper we proposed a framework named PeRView
which is meant for supporting selection of high quality personalized reviews
based on using a sub-set of micro-reviews which consistency more accurate than
the reviews based the proposed selection algorithm (\textit{MRS}).
The methodology used for review selection process exploits
a sub-set of micro reviews that highly covers the
whole domain of the micro-reviews set in order to
validate and select the best personalized reviews.
Since the select sub-set of micro-reviews are short descriptions,
they are used to match with reviews and based
on the maximum personalized coverage of
the selected sub-set of micro reviews in the reviews,
selection decisions are made based on the proposed evaluation metric.
In order to find the personalized similarity between a sentence in review and micro review,
three kinds of personalized similarity measures are defined and merged.
They are known as syntactical similarity which is based on traditional \textit{TF/IDF},
semantic similarity and sentiment similarity based on sentiment polarities.
Basically, the evaluation metric is designed based on these personalized similarities
as well as the size of the selected reviews
to make the final decision for selection the best reviews set.
Personalized user preferences are accommodated in the framework
to have more qualitative reviews. The experimental results show
that the proposed system is able to select
the best personalized reviews based on the toughest threshold value
(personalized coverage score) which is a 100\% accuracy.

\newpage

\bibliographystyle{plainnat}
\bibliography{mohammedref-PersonalRevSelection}

\end{document}